\documentclass[sigplan,screen]{acmart}
\setcopyright{cc}
\setcctype{by}
\acmDOI{10.1145/3774934.3786451}
\acmYear{2026}
\copyrightyear{2026}
\acmISBN{979-8-4007-2310-0/2026/01}
\acmConference[PPoPP '26]{Proceedings of the 31st ACM SIGPLAN Annual Symposium on Principles and Practice of Parallel Programming}{January 31 -- February 4, 2026}{Sydney, NSW, Australia}
\acmBooktitle{Proceedings of the 31st ACM SIGPLAN Annual Symposium on Principles and Practice of Parallel Programming (PPoPP '26), January 31 -- February 4, 2026, Sydney, NSW, Australia}
\received{2025-09-01}
\received[accepted]{2025-11-10}

\startPage{1}

\usepackage{booktabs}   
\usepackage{subcaption} 
\usepackage{soul}
\usepackage{listings}
\usepackage{colortbl}
\usepackage{microtype}
\usepackage{balance}

\graphicspath{{./fig}}

\begin{document}



\title[Rethinking Thread Scheduling Under Oversubscription: A User-Space Framework...]{Rethinking Thread Scheduling under Oversubscription: A User-Space Framework for Coordinating Multi-runtime and Multi-process Workloads}

\author{Aleix Roca}
\orcid{0000-0002-6715-3605}
\affiliation{%
  \institution{Barcelona Supercomputing Center}
  \city{Barcelona}
  \country{Spain}
}
\email{arocanon@bsc.es}

\author{Vicenç Beltran}
\orcid{0000-0002-3580-9630}
\affiliation{%
  \institution{Barcelona Supercomputing Center}
  \city{Barcelona}
  \country{Spain}
}
\email{vbeltran@bsc.es}

\begin{abstract}
The convergence of high-performance computing (HPC) and artificial intelligence (AI) is driving the emergence of increasingly complex parallel applications and workloads. These workloads often combine multiple parallel runtimes within the same application or across co-located jobs, creating scheduling demands that place significant stress on traditional OS schedulers. When oversubscribed (there are more ready threads than cores), OS schedulers rely on periodic preemptions to multiplex cores, often introducing interference that may degrade performance. In this paper, we present: (1) The User-space Scheduling Framework (USF), a novel seamless process scheduling framework completely implemented in user-space. USF enables users to implement their own process scheduling algorithms without requiring special permissions. We evaluate USF with its default cooperative policy, (2) SCHED\_COOP, designed to reduce interference by switching threads only upon blocking. This approach mitigates well-known issues such as Lock-Holder Preemption (LHP), Lock-Waiter Preemption (LWP), and scalability collapse. We implement USF and SCHED\_COOP by extending the GNU C library with the nOS-V runtime, enabling seamless coordination across multiple runtimes (e.g., OpenMP) without requiring invasive application changes. Evaluations show gains up to 2.4x in oversubscribed multi-process scenarios, including nested BLAS workloads, multi-process PyTorch inference with LLaMA-3, and Molecular Dynamics (MD) simulations.

\end{abstract}

\ccsdesc[500]{Software and its engineering~Multiprocessing / multiprogramming / multitasking}
\ccsdesc[500]{Software and its engineering~Scheduling}
\ccsdesc[300]{Software and its engineering~Application specific development environments}

\keywords{High Performance Computing, Thread Oversubscription, Runtime Systems, User-Space Scheduling}

\maketitle

\section{Introduction}
\label{sec:intro}



The convergence of HPC, data analytics, and artificial intelligence (AI) is transforming HPC workloads~\cite{ai_hpc_1,ai_hpc_2,ai_hpc_3,ai_hpc_4,ai_hpc_5}. Instead of monolithic applications, modern HPC workflows integrate multiple libraries and parallel programming models. This has significantly increased the complexity of execution environments, where multiple parallel applications may coexist and interact, leading to severe scheduling challenges.


HPC applications have traditionally relied on runtime systems to schedule parallel workloads on increasingly complex compute nodes with large multicore processors, NUMA architectures, and accelerators~\cite{hpc1,hpc2,hpc3,hpc4,hpc5}. These runtime systems are responsible for mapping application parallelism (parallel loops and tasks) to the underlying hardware while improving performance and mitigating bottlenecks. Runtimes typically leverage low-latency synchronization mechanisms such as busy-wait barriers and create one thread per core to avoid oversubscription.

Under oversubscription, the OS scheduler enforces fairness with time-sharing policies, where threads are preempted based on fixed time quanta. While controlled oversubscription has been shown to be beneficial in some cases~\cite{oversub_cloud,oversub_smp,oversub_mpi}, the OS lack of application awareness, generally leads to scheduling noise, disrupts critical-path computations, and pollutes caches~\cite{lhp,scalcollaps}.

Runtime composition poses a significant challenge in HPC~\cite{bolt,argobots,compo1}. When multiple runtime systems are used within the same process, they each create independent worker pools, leading to thread redundancy and resource contention. Moreover, since runtimes assume exclusive system access and use aggressive synchronization techniques, preemptions severely degrades their performance. In consequence, developers have historically avoided combining runtimes--not because it lacks utility, but because it typically causes oversubscription and poor performance. However, as HPC applications grow more complex and integrate heterogeneous software stacks, strict component isolation is becoming impractical.

Similarly, HPC applications are often run in exclusive mode to avoid interferences from other workloads and ensure predictable and efficient execution. Despite this, many applications exhibit varying degrees of parallelism during execution. These workloads transition between parallel phases and sequential phases, where underutilization of system resources occurs due to inherent algorithmic constraints. However, freely coexecuting processes in the same node to ``fill the gaps'' is discouraged because of the detrimental effects of oversubscription.




Existing solutions to mitigate oversubscription generally rely on resource partitioning~\cite{colo1,colo2,colo3,colo4} or using a single unified runtime for all involved components~\cite{python_composition,dlb,arachne}. However, the vast heterogeneity of HPC software stacks makes broad adoption of these solutions difficult, as competing APIs introduce additional complexity. Given these challenges, we believe that a transparent approach is necessary to facilitate adoption in the HPC community.



In this paper, we propose USF, a User-space Scheduling Framework with multi-process support that enables the creation of ad-hoc user scheduling policies. Unlike similar frameworks~\cite{arachne,dlb,ghost,syrup,f4}, USF requires no application modifications, no special permissions, works on vanilla Linux kernels, operates entirely in user space, and seamlessly supports Thread-Local Storage (TLS). We leverage USF to implement SCHED\_COOP, a seamless cooperative policy designed to mitigate oversubscription in HPC environments. SCHED\_COOP builds on the basis that HPC applications benefit from long uninterrupted runs with minimum interference. Applications scheduled by SCHED\_COOP are not periodically preempted, but instead are only swapped upon blocking. We implement both USF and SCHED\_COOP completely in user-space by extending the GNU C library and, in particular, its pthread API. Our solution supports any application that relies on the pthread synchronization APIs provided by glibc, covering the vast majority of runtime systems---such OpenMP (gomp, libomp), pthreadpool (PyTorch), or Rust.


Our evaluation demonstrates that (1) customized seamless user-space schedulers have potential to improve current scheduling limits, and (2) SCHED\_COOP is effective at mitigating oversubscription issues while preserving available parallelism. By minimizing thread preemptions and context-switching overhead, our policy improves application throughput and system efficiency. Since our framework operates on top of the Linux scheduler, critical OS services and non-cooperative threads remain unaffected.


%

The main contributions of this paper include:
\begin{itemize}
   \item USF, a seamless user-space scheduling framework that allows users to create their own scheduling policies.
   \item SCHED\_COOP, a USF cooperative scheduling policy to address oversubscription in HPC and AI workloads.
    \item A detailed evaluation in multi-runtime and multi-process execution scenarios.
    \item A practical approach to runtime nesting that avoids severe performance penalties.
    \item Guidelines for tuning oversubscribed applications.
\end{itemize}

\section{Background}
\label{sec:background}
This section details the main components of USF and SCHED\_COOP, implemented by extending glibc with the nOS-V~\cite{nosv} threading and tasking library.

\subsection{The Linux Kernel Process Scheduler}
\label{sec:os_sched}
The Linux kernel process scheduler is the ultimate authority in deciding which threads run on which cores. The scheduler is implemented in layered classes with increasing priority levels~\cite{sched_classes}. The default class is the Earliest Eligible Virtual Deadline First (EEVDF) scheduler~\cite{eevdf2,eevdf}. This policy attempts to schedule ready threads fairly, assigning each a quantum and preempting them if oversubscribed. Preemptive schedulers, such as EEVDF, can guarantee progress, but because preemptions happen without prior notice, they might interrupt critical operations and harm the performance of sensitive applications. The Linux scheduler exposes few configuration options to user-space, which limits the user’s ability to influence scheduling decisions.


\subsection{The GNU C Library}
\label{sec:glibc}
The GNU C Library~\cite{glibc} is one of the most common libraries in the Linux ecosystem. glibc provides the C runtime, the Linux dynamic loader, core C APIs (pthreads, I/O, math, etc.), and syscall wrappers. Its backward compatibility makes it the default choice for most compilers, which link it to all C applications by default. Therefore, glibc is the ideal convergence point to intercept process initialization, thread creation, blocking operations, affinity management, and, at the same time, transparently cover a wide range of applications. We implemented USF and SCHED\_COOP inside glibc and named it glibcv.

\subsection{nOS-V}
\label{sec:nosv}


nOS-V~\cite{nosv} is a flexible threading and tasking library with multi-process support that we use at the core of glibcv as detailed in Section~\ref{sec:design}. While nOS-V is fully documented in its original paper, we briefly summarize the key details here.

nOS-V provides bare-bones services for runtime systems. The nOS-V library defines the concept of task and its API provides the means to schedule and execute them in worker threads, which are managed by nOS-V transparently. At a glance, the nOS-V API allows to create tasks, submit them to the scheduler, run callbacks upon task completion, pause tasks, and resume them. By leveraging these fundamental functionalities, it is possible to build a full parallel runtime system with dependency management and standard features (e.g. taskwait), such as NODES~\cite{nodes} (task-based) or libompv~\cite{ompv} (task and fork-join). Hence, nOS-V can be seen as a library to help build runtimes.


nOS-V initializes a worker pool and always keeps exactly one running worker pinned per core at all times. Idle workers wait until the centralized scheduler assigns them a task, which they run to completion. However, tasks can suspend themselves while waiting on user-defined conditions (e.g., synchronization points like \texttt{taskwait}). When this happens, nOS-V performs a worker swap, replacing the suspending worker with another to keep the core busy. Once the wait condition is met, the task is requeued into the scheduler and later assigned to an idle worker. If a worker is serviced a task already bound to a different worker, nOS-V performs another swap so both the original worker and its task resume on the newly allocated core. nOS-V provides several algorithms to select the next task to run; we detail the strategy used in SCHED\_COOP in Section~\ref{sec:td1}.

A key nOS-V feature is seamless multi-process scheduling. Internally, nOS-V relies on a shared memory segment~\cite{shm} to keep tasks, workers, and other relevant structures. Each process creates its own tasks, but a single instance of the centralized scheduler manages all tasks from all processes. Idle workers can be handed tasks belonging to any process but only execute their own. If a worker receives a task belonging to another process, the worker swaps itself with a worker of that process and transfers its task to it. nOS-V allocates a configurable quantum per-process but never preempts involuntarily. Instead, process switches occur only at scheduling points (task suspension, end, or yield).

\section{SCHED\_COOP}
\label{sec:sched_coop}



From the user’s perspective, SCHED\_COOP resembles a Linux scheduling policy like SCHED\_RR (real-time round-robin). Unlike those, however, it is a \textbf{virtual} scheduler implemented entirely in user space via USF.

Processes enter the SCHED\_COOP at launch time with the new ``\lstinline{chrt -c <app>}'' command option. Child processes and threads inherit it by default at runtime, and only exit it at termination time. SCHED\_COOP threads run uninterruptedly with a fixed single-core affinity until the application conditions (not OS conditions) compel it to wait. Only when a thread blocks or voluntarily relinquishes its core, another thread might be resumed on it.

Although SCHED\_COOP threads run greedily, they do not stall non-SCHED\_COOP processes, since the default OS scheduler may still preempt them. The key distinction is that SCHED\_COOP threads never preempt one another. In this sense, SCHED\_COOP resembles SCHED\_RR, where threads run until they yield or block. However, unlike SCHED\_COOP, SCHED\_RR threads cannot be preempted by regular threads (only by higher-priority SCHED\_RR threads), which risks stalling the system; hence SCHED\_RR is restricted to root and system users. By contrast, SCHED\_COOP is available to regular users and, at worst, its threads may stall only other SCHED\_COOP threads.

The SCHED\_COOP goal is to improve throughput and system efficiency by improving scheduling decisions in oversubscription scenarios. Therefore, it is not intended for all workloads, but for parallel applications or multi-process workloads leveraging multiple runtime systems prone to oversubscription.


\section{Design}
\label{sec:design}

This section follows a three-layered top-down approach. Section~\ref{sec:td1} explains how USF and SCHED\_COOP operate from a threading perspective. Section~\ref{sec:glibcv} describes how this behavior is achieved by coupling glibc with nOS-V. Section~\ref{sec:api} delves deeper into the extended glibc features and the nOS-V API usage, and Section~\ref{sec:limitations} discusses this implementation limitations.


\subsection{USF and SCHED\_COOP Threading Overview}
\label{sec:td1}

USF works seamlessly by extending standard glibc APIs. At a glance, it manipulates thread creation, destruction, and blocking APIs. Pthread creation is extended to prevent threads from running freely. Instead, they are blocked and queued until an idle core is available. A centralized scheduler monitors available queued threads and dispatches them to idle cores according to the current policy. Threads resume their execution pinned to an idle core and run uninterruptedly until they end, block, or yield. When a thread reaches a blocking operation (e.g., a contended \texttt{pthread\_mutex\_lock}), it notifies the scheduler that its core is about to become idle. In response, the scheduler pins and resumes another thread on the idling core. Unblocking paths (e.g., \texttt{pthread\_mutex\_unlock}) notify the scheduler of threads ready to be resumed, but these threads are not resumed immediately. Instead, they are queued within the scheduler.

SCHED\_COOP is a policy that defines a simple algorithm to select the next thread to run. By default, all threads have a preferred affinity set to the last core on which they ran. Once a previously blocked thread becomes ready, it is queued in a per-process, per-core First-In First-Out (FIFO) queue. The scheduler first assigns threads to idle cores matching their affinity; if none are available, it searches within the same NUMA domain, and otherwise assigns them to any other NUMA node. When the current process' quantum expires (20 ms by default), evaluated at task scheduling points, the scheduler continues assigning tasks from the next process’s queues. The scheduling logic is part of nOS-V, as explained in Section~\ref{sec:nosv}.

%

\subsection{glibcv: Coupling glibc and nOS-V}
\label{sec:glibcv}

\begin{figure}
    \centering
    \includegraphics[width=\columnwidth]{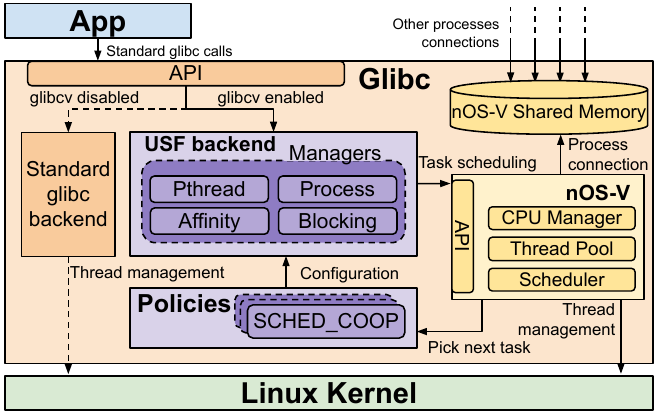}
    \caption{Glibcv architecture diagram. Application's standard API calls are forwarded to the USF backend if enabled, which bridges with the nOS-V API. nOS-V schedules threads according to the selected policy.}
    \Description{Layered USF architecture diagram comparing two execution paths between an application and the Linux kernel. A standard glibc call from an application can be routed through the standard glibc API as usual, or it can be routed through the USF backend. The USF backend features four subsystems to manage parallelism: pthread management, affinity management, process management, and blocking APIs management. USF connects with nOS-V runtime, and nOS-V uses kernel APIs to schedule threads. nOS-V connects with a shared memory alongside other processes to manage multi-process execution. Next to the USF and the nOS-V block, the Policies block shows a stack of policies, featuring SCHED\_COOP, that communicates with both nOS-V to select the next task and USF for configuration.}
    \label{fig:arch}
\end{figure}

Figure~\ref{fig:arch} shows the glibcv architecture diagram. Both USF and SCHED\_COOP are implemented within glibc, and they both rely on the nOS-V threading library to manage most of the scheduler logic described before. Conceptually, we use nOS-V to convert glibc into a task-based runtime; all pthreads become worker threads with an associated task, and most synchronization APIs become task block/resume operations. Since each task remains bound to the same worker thread, they are fully compatible with TLS.

When a pthread is created, glibcv transparently converts it into a nOS-V worker thread and assigns it a new task mapped to the pthread's entry function. The task is then submitted to the nOS-V scheduler and later resumed on an idle core with a fixed single-core affinity, where it runs until the user code ends, blocks, or yields. When the user code ends, the task ends, and nOS-V schedules another worker in its core. On blocking calls such as \texttt{pthread\_mutex\_lock}, the task suspends and nOS-V also schedules another worker in that core. On unblocking calls, such as \texttt{pthread\_mutex\_unlock}, the suspended task is resubmitted to the scheduler. Forked processes and their threads are also registered within nOS-V, which manages all tasks in a single multi-process centralized scheduler.

Note that because we prevent threads from running freely and nOS-V carefully keeps a single thread per core, threads must run in an orderly manner, and we avoid preemptions between USF threads. See Section~\ref{sec:nosv} for more details on nOS-V scheduling.

\subsection{Extended API Calls}
\label{sec:api}

In this section, we explain the implementation details of the extended glibc APIs with nOS-V, as depicted in the Figure~\ref{fig:arch} USF backend block.

\subsubsection{Pthread Creation}
\label{sec:pthread}

Whenever \texttt{pthread\_create()} is called, glibcv swaps the user-provided function pointer with a wrapper. Once the thread is created, the wrapper calls the \texttt{nosv\_attach()} API, which registers the calling thread as a nOS-V worker, assigns the thread's entry function to a new nOS-V task, assigns the task to the worker, submits the task to the scheduler, and immediately blocks the current thread. At this point, glibcv has created a pthread following the provided user attributes (except for the affinity), but it has recruited it as a nOS-V worker, which means that it can no longer run freely. We have extended the pthread\_t object to accommodate the nOS-V task object and related data.


USF uses a thread caching strategy, similar to that proposed by D. Dice and A. Kogan~\cite{caching}, to avoid the overhead of repeatedly creating and destroying threads. When a thread’s user function ends (or calls \texttt{pthread\_exit()}, unwinding the stack), the thread blocks and is swapped out with another worker instead of exiting. If another thread calls \texttt{pthread\_join}, the join operation is masked and, instead, the completed thread is placed in the cache. Subsequent \texttt{pthread\_create} calls reuse the most recent cached threads rather than creating new ones. At shutdown, all cached threads are unblocked, detached, destroyed, and truly joined. Threads are detached by calling \texttt{nosv\_detach()}, which deregisters them from nOS-V and terminates their task.

\subsubsection{Affinity Management}

Attempts to change thread affinity (e.g., \texttt{pthread\_setaffinity\_np()}) under USF are considered hints. This is necessary to prevent interfering with the nOS-V fine thread placement. In such cases, glibcv stores the user affinity in the \texttt{pthread\_t} object, but does not change it. To preserve application compatibility, when queried with \texttt{pthread\_getaffinity\_np()} and similar APIs, glibcv returns the previously stored affinity instead of the real affinity. For the same purpose, we have also extended the \texttt{syscall()} system call entry point to intercept \texttt{sched\_setaffinity}, which is often called directly because it has different semantics than the glibc wrapper. We also use hash tables to keep track of tid-task mapping, necessary for the \texttt{sched\_set/getaffinity} calls. However, if needed, it would be simple to create a USF policy that respects the user affinity.

\subsubsection{Forks and Process Initialization}

Glibcv can manage multiple processes. At process startup, if the \texttt{USF\_ENABLE} environment variable is set, glibcv initializes nOS-V before any constructor is called. As explained in Section~\ref{sec:nosv}, nOS-V connects to any pre-existing shared memory segment and registers the new process. Then, glibcv converts the current main thread into a nOS-V worker with an attached task using \texttt{nosv\_attach()}. When the process terminates or when exec is called, glibcv calls \texttt{nosv\_shutdown()} to deregister the process and shutdown its worker threads. In the case of exec, after the operation completes, the new application also finds the \texttt{USF\_ENABLE} environment variable and registers the process into nOS-V again.

\subsubsection{Blocking APIs}

\begin{figure}
\begin{lstlisting}[language=C, caption={Glibcv mutex lock/unlock extension. When a mutex is contended, threads are placed in a spinlock-protected per-mutex FIFO wait queue before blocking in nOS-V. On unlock, if the queue is not empty, one thread is dequeued and submitted to the scheduler.}, label=lst:mutex]
int pthread_mutex_lock(
      pthread_mutex_t *mutex) {
  nosv_task_t task = pthread_self()->task;
  pthread_spin_lock(&mutex->spinlock);
  if (pthread_mutex_trylock(mutex)) {
    //Contended
    queue_add_tail(&mutex->queue, task);
    pthread_spin_unlock(&mutex->spinlock);
    nosv_pause();
  } else {
    pthread_spin_unlock(&mutex->spinlock);
  }
  return 0;
}
int pthread_mutex_unlock(
      pthread_mutex_t *mutex) {
  pthread_spin_lock(&mutex->spinlock);
  list_head_t wait = &mutex->queue;
  nosv_task_t task = queue_pop_head(wait);
  if (!task)
    //Standard mutex unlock
    __pthread_mutex_unlock(mutex);
  pthread_spin_unlock(&mutex->spinlock);
  if (task)
    nosv_submit(task);
  return 0;
}
\end{lstlisting}
\Description{The listing shows two functions corresponding to extended pthread\_mutex\_lock and pthread\_mutex\_unlock.}
\end{figure}

We extended the mutex, condition variable, barrier, semaphore, sleep, yield, and polling (poll, epoll) APIs to detect blocking operations. Listing~\ref{lst:mutex} shows the mutex pseudocode. We augmented \texttt{pthread\_mutex\_t} with a FIFO wait queue and a spinlock to protect it. On contention, \texttt{pthread\_mutex\_lock()} enqueues the calling thread’s nOS-V task (each pthread is associated with one), and blocks it via \texttt{nosv\_pause()}, which triggers a worker swap. On \texttt{pthread\_mutex\_unlock}, if contended tasks exist, one is submitted to the nOS-V scheduler (\texttt{nosv\_submit()}) and ownership is transferred to that task; only if no tasks are queued is the mutex fully released.

Other APIs follow the same pattern: contended tasks are tracked in standard glibc objects and managed with nOS-V calls to block, resume, yield, or wait. Timed variants of poll and epoll periodically perform non-blocking event checks using the original API, then call \texttt{nosv\_waitfor()} to perform a short timed wait in a loop (5ms by default) until either the user timeout expires or an event occurs. \texttt{nosv\_waitfor()} also triggers a worker swap, but tasks are automatically resubmitted once the provided timeout elapses.


\subsection{Limitations}
\label{sec:limitations}

USF efficient use of cores partly depends on its capacity to intercept wait operations, such as blocking operations. If a wait operation is not intercepted, nOS-V is unable to detect the idle core, and it cannot run other tasks on it until the operation completes. Instrumenting blocking operations is simple because these are likely to use standard glibc APIs or a kernel syscall wrapper. However, some applications or runtimes rely on busy-waiting barriers to implement low-latency synchronizations. These cases are complex to instrument at the glibc level, but failing to do so might stall progress and deadlock the application. This can occur if the number of waiting threads exceeds the number of cores or if multiple barriers are active simultaneously. Instead, preemptive schedulers mask this into a performance problem because they do not depend on scheduling points to guarantee progress in oversubscribed systems. As we further detail in Section~\ref{sec:interference}, our solution to these cases is to manually adapt the busy-wait barriers to include an occasional yield operation (such as \texttt{sched\_yield()}) every few busy iterations, which is not only beneficial for policies such as SCHED\_COOP, but also for oversubscribed preemptive schedulers (see Section~\ref{sec:matmul}).



USF supports any application that relies on the glibc synchronization APIs. Most runtimes typically offer a POSIX-compatible backend alongside platform-specific alternatives for Windows, macOS, or Linux (e.g., custom futex-based implementations). Extending USF to support futex-based backends via glibc’s futex wrapper is part of our planned future work.

USF policies inherit the nOS-V security limitations. nOS-V shares state via a shared memory segment. This shared memory is visible to all processes connected to the same nOS-V instance, and sensitive information could be exposed between them. For this reason, we only allow processes of the same user and group to connect to the same shared memory. In the future, we will study a tighter nOS-V integration with the Linux kernel to overcome this limitation.

Because glibcv extends standard glibc objects that are statically allocated (such as \texttt{pthread\_mutex\_t}) and its size is calculated at compile-time, all applications and their dependencies must be rebuilt against glibcv. Fortunately, build systems such as Nix~\cite{nix} are capable of automating whole system builds, which greatly simplifies testing glibcv with applications with complex dependencies, such as PyTorch.

\section{Evaluation}
\label{sec:evaluation}

In this section, we evaluate USF and SCHED\_COOP across progressively more complex scenarios. We begin by outlining our evaluation methodology, which follows a bottom-up approach. Our first hypothesis is that a user-space scheduler, combined with manual adaptations to the software stack, can improve the performance of oversubscribed applications. Our second hypothesis is that similar gains can be achieved seamlessly, without invasive modifications. To validate these hypotheses, Section~\ref{sec:methodology} introduces the evaluation scenarios---nested runtimes and multi-process coexecution---and describes the experimental setup. Then, Section~\ref{sec:interference} addresses interference considerations, ensuring fair comparisons by harmonizing runtime wait policies and disabling thread pinning. Section~\ref{sec:matmul} validates both hypotheses with a controlled matrix multiplication benchmark, contrasting manual adaptations with the seamless USF approach. Section~\ref{sec:cholesky} extends the analysis to a Cholesky factorization benchmark with multiple runtime compositions, highlighting the challenges of heterogeneous software stacks. Next, Sections~\ref{sec:ms} and \ref{sec:lmp} demonstrate USF in a practical AI microservices and molecular dynamics scenarios, respectively, showcasing its benefits in realistic multi-process environments.

\subsection{Evaluation Scenarios and Setup}
\label{sec:methodology}


\begin{table}[]
\caption{Evaluation machine specifications.}
\label{tab:machines_spec}
\centering
\begin{tabular}{|c|c|}
\hline
\textbf{Machine}        &Marenostrum 5               \\ \hline
\textbf{Memory}         &256GiB                      \\ \cline{1-1}
\textbf{Network}        &InfiniBand (100Gb/s)        \\ \cline{1-1}
\textbf{CPU Model}      &Intel Sapphire Rapids 8480+ \\ \cline{1-1}
\textbf{GPU}            &-                           \\ \cline{1-1}
\textbf{CPU Arch.}      &x86\_64                     \\ \cline{1-1}
\textbf{CPU Cores}      &56x2                        \\ \cline{1-1}
\textbf{L1, L2, L3}     &80KB, 2MB, 105MB            \\ \hline
\end{tabular}
\end{table}

We concentrate on two representative oversubscription scenarios within a single node: nested runtimes, where threads from an outer runtime invoke parallel regions in an inner runtime, causing the number of active threads to grow rapidly (e.g., an application with $n$ oneTBB threads each calling a BLAS routine that spawns an OpenMP region with $n$ threads, yielding up to $n^2$ threads); and multi-process workloads, where several parallel applications overlap sequential and parallel phases to maximize resource use.
Our experiments compare performance under the standard Linux scheduler (baseline) and SCHED\_COOP. We use the production system described in Table~\ref{tab:machines_spec}, where kernel modifications are not allowed, demonstrating the value of our transparent user-space approach. Our software environment includes: Linux 5.14, glibc 2.40-6, nixpkgs 25.05, clang 20, gcc 14.2.1, OpenBLAS 0.3.28~\cite{openblas}, BLIS 1.1~\cite{blis1,blis3}, PyTorch 2.6~\cite{pytorch}, Transformers 4.51.3~\cite{transformers}, oneTBB 2020.3~\cite{tbb}, MPICH 4.3.0~\cite{mpich}, LAMMPS 29Aug2024\_update2~\cite{lmp1,lmp2}, DeePMD-kit 3.1.1~\cite{dmd1,dmd2,dmd3}, nOS-V 3.2~\cite{nosv}, Nanos6 4.2~\cite{nanos6}, and NODES 1.3~\cite{nodes}.

\subsection{Interference Considerations}
\label{sec:interference}


In order to achieve a fair comparison, we first consider reasonable parameters to runtimes and other libraries that would otherwise clutter the results concerning the differences between the Linux process scheduler and SCHED\_COOP. These changes are necessary because most runtimes and some parallel libraries are not designed to coexist in the same address space. We have tested our applications with and without these adaptations and decided to ultimately include them because of their performance benefits for both the baseline and SCHED\_COOP schedulers.


Wait policies are common to most runtimes, such as OpenMP (OMP\_WAIT\_POLICY) and OmpSs-2~\cite{ompss2}. They define the behavior of worker threads upon work starvation. By default, both runtimes assume a hybrid wait policy where workers embrace a timed active wait prior to taking the expensive OS block/resume path. However, a runtime worker actively waiting for work might be wasting CPU cycles and delaying the execution of another runtime worker with ready tasks to execute. In consequence, all of our experiments have their runtimes configured with similar "passive" wait policies (recommended under oversubscription), in which threads immediately take the OS blocking path instead of actively waiting for more work. Thread pinning is another feature that lets users configure if and how worker threads are pinned to which system cores (e.g., OMP\_PLACES for OpenMP). However, our early evaluation showed that multiple runtimes or oversubscribed processes pinning their workers to overlapping CPU sets interfere with each other, dramatically hindering performance. All our tests avoid these problems by configuring runtimes to skip thread pinning.

Expanding our analysis beyond runtime systems, we observed a recurring pattern: some libraries implement custom busy-waiting barriers instead of relying on standard APIs. In systems with no more active threads than available cores, busy-waiting barriers can be an effective way to briefly stall threads. However, in oversubscribed systems, scheduling noise can significantly delay barrier completion. A single preempted thread can block an entire team, and worse, busy-waiting threads continue consuming CPU cycles until their time slice expires---potentially delaying the very thread needed to release the barrier. This issue is further exacerbated when multiple busy-waiting barriers operate simultaneously, a scenario frequently encountered when parallel BLAS kernels execute concurrently. For our evaluation, we follow a common pattern: to yield within the busy-wait barriers to allow other threads to progress. We do so by modifying a single line of code in OpenBLAS, BLIS, and MPICH, to add the \texttt{sched\_yield()} syscall (details in Section~\ref{sec:matmul}).

A side effect of nesting parallel runtimes is that the resulting kernel granularity might be too small. A nested parallel loop or similar structure gets subdivided into an exponential number of micro-kernels. Fine-grained kernels on runtime systems pose a challenge on its own~\cite{cilk,coro,xopenmp,taskgraph} because the runtime and system overheads become critical. In this study, however, we focus on the analysis of oversubscription, and we assume sufficiently large workloads.

\subsection{Manual and Seamless Scheduling Comparison}
\label{sec:matmul}

In this section we evaluate our two hypotheses as defined in Section~\ref{sec:methodology} using a simple Matrix Multiplication (matmul) benchmark. Our matmul (see Listing~\ref{lst:matmul}) is an example of the nested runtimes case, where we applied blocking to subdivide the problem with OmpSs-2 tasks (outer runtime) and process each block using BLIS parallelized with LLVM's OpenMP (inner runtime). Each OmpSs-2 task is executed by a Nanos6 thread, and each task opens an OpenMP parallel region within the BLAS library.


\begin{figure}
\begin{lstlisting}[language=C, caption={Simplified matmul code parallelized with runtime nesting: OmpSs-2 (Nanos6) and BLIS (OpenMP). The problem is subdivided in NB blocks, and each block is of size TS. The "oss task" pragma creates a task with the specified dependencies, and the "oss taskwait" blocks execution until all previous tasks have been completed.}, label=lst:matmul]
for (k = 0; k < NB; k++)
  for (i = 0; i < NB; i++)
    for (j = 0; j < NB; j++)
      #pragma oss task inout(C[i][j]) \
        in(A[i][k]) in(B[k][j])
      cblas_dgemm(TS,
        A[i][k], B[k][j], C[i][j]);
#pragma oss taskwait
\end{lstlisting}
\Description{matmul user code featuring a blocked layout with three for loops. Inside the inner loop, a task is created with pragma omp task. The task calls the clbas_dgemm API to compute the block.}
\end{figure}


We fix all input matrix sizes to $32768\times32768$ and evaluate the benchmark response to oversubscription by varying the task size (outer runtime parallelism) and number of OpenMP threads (inner runtime parallelism). For example, in our 56-core system, with 2 OpenMP threads and 16384 task size (at most $(32768/16384)^2=4$ parallel tasks at the same time), we have a total of $2\times4=8$ threads busy at the same time and the system is mostly underused. With 56 OpenMP threads and 512 task size we have 229376 busy threads and the system is heavily oversubscribed. We run each kernel in a loop for at least 60s and measure MOPS/s performance metric computed as $\frac{size \cdot loops}{seconds}\cdot 10^{-6}$.

\begin{figure}
    \centering
    \includegraphics[width=1\linewidth]{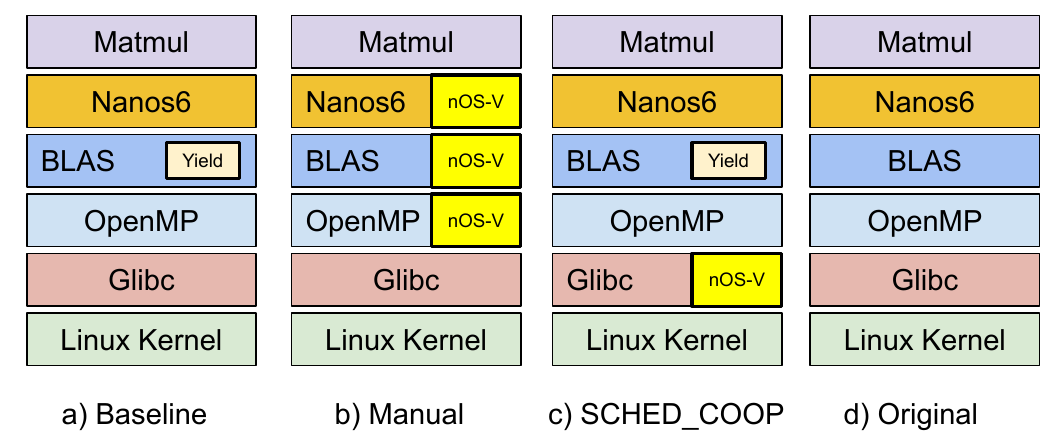}
    \caption{Evaluated matmul software stacks. a) Baseline with yield. b) Manual nOS-V integration. c) Seamless nOS-V integration. d) Unmodified (no yield). }
    \Description{The figure shows four software stacks labeled with letters from "a" to "d". All stacks have the same layers, but exhibit modifications between them. The layers are, from bottom to top: Linux Kernel, glibc, OpenMP, BLAS, Nanos6, and Matmul. The stack "a" show the Baseline stack, where the BLAS layer is modified with yield. The "b" stack shows "Manual", where Nanos6, BLAS, and OpenMP has been modified with nOS-V. Ths stack "c" shows the SCHED\_COOP stack, which has a modified BLAS with yield, and a modified glibc with nOS-V. The "d" stack shows the Original version, that has not been modified.}
    \label{fig:matmul_stacks}
\end{figure}



We compare four matmul versions, shown in Figure~\ref{fig:matmul_stacks}, ordered from least to most invasive changes: Original is the unmodified implementation. Baseline is identical to Original but incorporates yielding on barriers (Section~\ref{sec:interference}) and serves as our reference point. SCHED\_COOP uses the same software stack as Baseline, except that glibc is replaced with glibcv to enable our scheduler through nOS-V. The Manual version implements the same scheduling policy as SCHED\_COOP, but instead of enabling it transparently through glibcv, it is integrated manually via explicit calls to the nOS-V threading library within the Nanos6, OpenMP, and BLAS libraries. The goal of the Manual configuration is to establish an upper bound for glibcv’s performance by maximizing gains without the constraints of a seamless solution, incorporating minor ad-hoc optimizations at each library level. In essence, Manual attaches pthreads created at the different levels with \texttt{nosv\_attach} and uses nOS-V blocking primitives instead of pthread primitives.


\begin{figure*}
    \centering
    \includegraphics[width=1\linewidth]{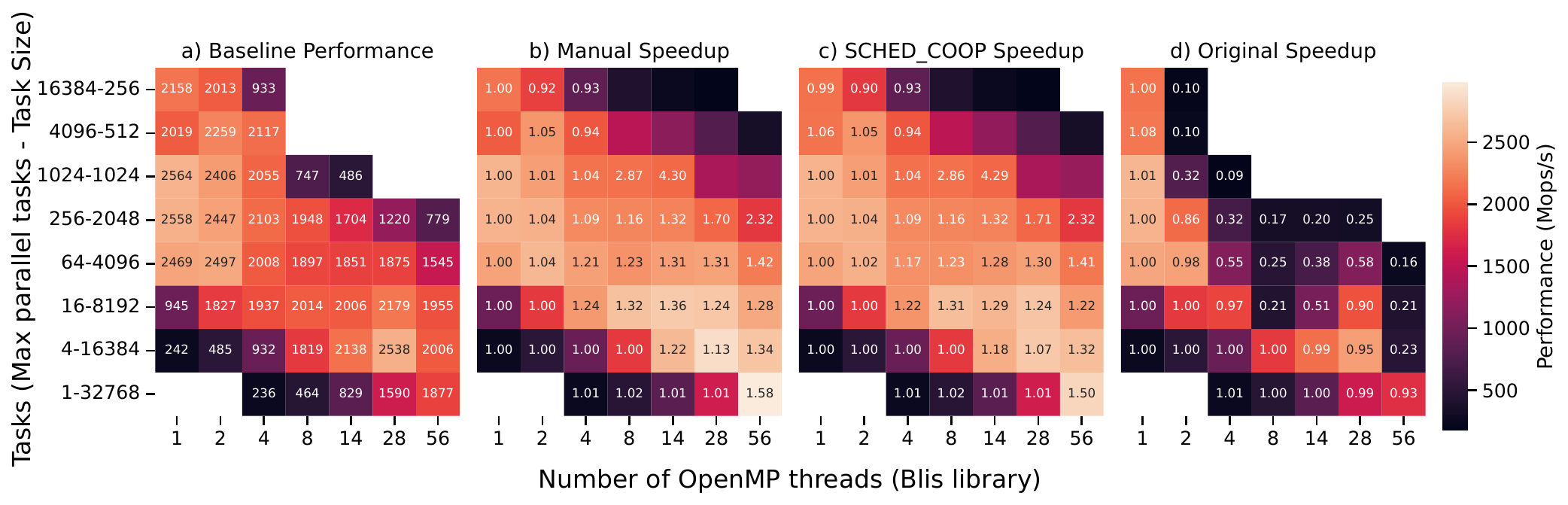}
\caption{Evaluation of matmul with two nested runtimes. Figure a) numbers show performance (higher is better), while b), c), and d) show element-wise speedup with respect to a). The colors of all heatmaps show performance, not speedup. White squares are timed out experiments (\textgreater 15 minutes). Squares without speedup lack the Baseline reference because of time out.}
    \Description{Four heatmaps are shown, from left to right: Baseline Performance, Manual Speedup, SCHED\_COOP speedup, and Original speedup. Each heatmap corresponds to a different software stack. Each software stack is evaluated by changing the number of OpenMP threads used in BLIS (x axis) and Nanos6 tasks (y axis). The x-axis range is which include 1,2,4,8,14,28, and 56. The y-axis range is 1,4,16,64,256,1024,4096, and 16384 maximum parallel tasks. The heatmaps show performance increase for Manual and SCHED\_COOP when compared to baseline in the top-left to bottom-right diagonal. Original obtains the worst performance.}
    \label{fig:matmul}
\end{figure*}

Figure~\ref{fig:matmul} shows the results of the experiment. For each evaluated matmul version, the heatmap colors helps us to roughly distinguish three areas: The lower left triangular area corresponds to under-used resource configurations (fewer threads than cores and coarse tasks). The top-right triangular areas represent overly fine-grained blocks caused by exponential fragmentation (more threads than cores and fine-grained tasks). The middle diagonal area corresponds to oversubscribed configurations with reasonable granularities. We consider the upper and lower triangular areas to be ill-designed: either not enough or too much parallelism is exposed. Instead, our focus is on the middle areas, which include a large range of valid configurations. Note that the configurations with a single runtime enabled, i.e. (1-32768, *) and (*, 1), consider scenarios without runtime composition.


As per our first hypothesis, we compare Baseline with Manual (see Figure~\ref{fig:matmul} a) and b), respectively) to demonstrate that the nOS-V user-space scheduler can improve the performance of an oversubscribed application. At first glance, the heatmap coloring reveals how the Manual version is capable of expanding the "hotter" (more performant) regions with respect to Baseline, particularly in the middle area. Then, the speedups depicted in Manual detail an improvement between 1\% and 36\% depending on the configuration. Hence, we can conclude that the first hypothesis is valid. Note that although we achieve up to 4x speedup on some configurations, we do not consider them relevant because the resulting performance is low when compared to the best results obtained.

As per our second hypothesis, we compare Baseline with SCHED\_COOP (see Figure~\ref{fig:matmul} a) and c), respectively) to demonstrate that a seamless approach can also improve the performance when under oversubscription. By using the same method, we can see a similar performance improvement by comparing the color gradients of both heatmaps and observe a speedup between 2-28\%. Thus, we consider our second hypothesis validated as well. Moreover, the best SCHED\_COOP configuration (28,4-16384), which involves the two runtimes, improves the best Baseline configuration (1, 1024-1024), with a single runtime, by almost 9.8\%. Also, the best Manual configuration (28,4-16384) improves the best Baseline by 11.77\%. This shows that mixing runtimes in the same address space can be beneficial if under the appropriate scheduling policy.


In this experiment, the matmul with nOS-V versions obtain a performance increase mostly because of how BLAS busy-wait barriers are managed. Essentially, even if \texttt{sched\_yield} is commonly used to relinquish the current core, Linux might not yield immediately but instead waits until the thread's quantum expires~\cite{oversub_mpi}. The details depend on the Linux version used, but its usage is still beneficial because threads yield as soon as possible instead of waiting for the next clock interrupt. For comparison, see the results of Original in Figure~\ref{fig:matmul} d). This version is the same as Baseline, but it uses the original busy-wait barrier without \texttt{sched\_yield}. We can see large slowdowns in many configurations, particularly the most oversubscribed because most time is spent busy-waiting. Instead, the matmul SCHED\_COOP version always yields because the original \texttt{sched\_yield} uses a nOS-V yield API internally. Yielding sooner is beneficial for this scenario because it allows threads to exit barriers earlier when under heavy oversubscription.


In conclusion, both Manual and SCHED\_COOP can improve the range of configurations that attain higher performance, with the advantage that SCHED\_COOP can do it transparently. USF enables the implementation of tailored user-space scheduling policies without kernel modifications, such as SCHED\_COOP. This has the potential to effectively unlock runtime composition and oversubscription scenarios, which so far had generally been avoided because of their penalties. It also eases the exploration of the input configuration space. Instead of searching for the right thread balance between runtimes, users can focus on finding configurations that expose enough parallelism while maintaining coarse-enough tasks.



\subsection{Runtime Compositions}
\label{sec:cholesky}


In this Section, we show how complex the runtime composition combinations can quickly become even if working with a relatively simple benchmark. To do so, this time we use a Cholesky factorization benchmark and multiple combinations of different runtime implementations, including GNU's gomp, LLVM's libomp, and Intel's tbb. Similar to matmul, our Cholesky first divides the problem into blocks using an outer runtime, then for each block, it calls a BLAS API that further subdivides and computes the kernel in parallel using an inner runtime (either OpenMP or pthread based). For brevity, we present results for a subset of relevant Baseline and SCHED\_COOP configurations. We fix the input size and task size to $32768\times32768$ and 1024 respectively. The results are shown in Table~\ref{tbl:cholesky}. SCHED\_COOP outperforms the baseline in all cases, particularly in the pth cases. That's because the BLIS pthread backend does not reuse threads but creates and destroys them for every dgemm call. Instead, USF caches threads transparently and prevents OS noise (see Section~\ref{sec:pthread}). We found that the pthread cache roughly multiplies by 4 the base SCHED\_COOP performance. On the other cases, the pthread cache had no effect because runtimes reuse pthreads efficiently. Note that the outer "gnu" and inner "llvm" use static builds to prevent symbol collision.

\begin{table}[]
\caption{Cholesky evaluation results combining multiple runtimes for three degrees of parallelism. For each degree, we report the performance in MOPS/s (higher is better) of Baseline and the SCHED\_COOP speedup. Mild uses 8x8 threads (1.14 threads per core), Medium uses 14x14 (3.5 threads per core), and High uses 28x28 (14 threads per core). gnu stands for GNU's OpenMP, llvm for LLVM's OpenMP, pth for the pthread BLAS backend, and opb for OpenBLAS.}
\label{tbl:cholesky}
\begin{tabular}{|c|c|c|c|c|c|}
\hline
\rowcolor[HTML]{EFEFEF} 
Out & Inn  & BLAS & Mild        & Medium     & High       \\ \hline
gnu & llvm & opb  & 1468, 1.1x  & 580, 2.36x & 347, 3.14x \\ \hline
tbb & llvm & opb  & 1619, 1.1x  & 602, 2.47x & 363, 3.22x \\ \hline
tbb & gnu  & blis & 2004, 1.06x & 694, 3.23x & 358, 4.28x \\ \hline
tbb & pth  & blis & 1590, 1.34x & 356, 6.28x & 103, 14.4x \\ \hline
gnu & pth  & blis & 1413, 1.33x & 254, 6.86x & 86, 14.7x  \\ \hline
\end{tabular}
\end{table}

%
%

The key takeaway is that even for simple applications, we might have multiple combinations of runtimes, implementations, and compilation options that can quickly complicate our strategies to handle oversubscription. Coordinating all these scenarios manually would entail a huge effort in understanding and adapting each possibility. Instead, SCHED\_COOP is capable of managing all these cases transparently, while keeping the user in control.



\subsection{Microservices Case Study}
\label{sec:ms}

This section examines a Python multi-process AI inference benchmark. Python dominates AI development, but its Global Interpreter Lock (GIL) serializes threads, prompting developers to use multiple processes with native parallel libraries instead. We leverage this common pattern to demonstrate SCHED\_COOP in an oversubscribed multi-process environment, comparing it against various resource partitioning schemes under the standard Linux scheduler. We focus on CPU-based inference, which is gaining research attention~\cite{iacpu1,iacpu2,iacpu3,iacpu4,iacpu5} due to CPUs' ubiquity, flexibility, and cost-effectiveness.

Our benchmark simulates periodic client requests following a Poisson distribution, processed by four Python server processes on a dedicated node. A Gateway process receives requests, simulates planning logic, and forwards them to three inference servers in parallel, blocking until all responses return. Each server runs a distinct model: Meta's LLaMA 3.2 (1B)\cite{llama}, OpenAI's GPT-2 (124M)\cite{gpt2}, and a fine-tuned Meta RoBERTa-large (355M)~\cite{roberta} for sentiment analysis. All models run on PyTorch, which uses efficient C++ libraries and OpenBLAS with OpenMP to bypass GIL limitations. Thus, each user request creates one thread per process (four total), with three threads spawning pthread teams for parallel inference computation. At high request frequencies, overlapping requests cause system oversubscription.

We evaluate the behavior of our benchmark as we increase the request frequency. Each execution generates 28 requests of 128 tokens ($\sim$168 words) and simulates a grouping of 8 batches per request. Each model outputs at most 32 tokens ($\sim$42 words). We do not limit the maximum number of outer threads servicing requests, but we fix the number of inner BLAS threads used to compute each inference. We calculated the ideal number of inner threads per model by running an isolated strong scalability test with a single request. In particular, LLaMA obtained 5.4s with 28 cores, GPT-2 1.8s with 8 cores, and RoBERTa 1.2s with 8 cores.

We compare the results of running the benchmark under SCHED\_COOP with four baseline scenarios. \texttt{bl-none} employs no resource partitioning. \texttt{bl-eq} splits equally its cores among all Python inference servers but for the Gateway, which uses 2 cores. \texttt{bl-opt} uses an optimized partitioning scheme based on the scalability results obtained earlier to proportionally assign cores to the most demanding servers minus two cores for the Gateway server. The optimized partitioning splits the node's 112 cores in 71 cores for LLaMA (~64\%), 23 cores for GPT-2 (~21\%), and 16 cores for RoBERTa (~14\%). Note that even with this resource partition, there is still oversubscription within each partition as the request frequency increases. However, partitioning helps limit overall oversubscription. \texttt{bl-none-seq} uses no partitioning and avoid oversubscription by computing inferences sequentially (without OpenMP parallelism).

All tests use a nice priority of 0 for the Gateway server (high priority) versus a priority of 20 (less priority) for the inference servers in order to attend and respond user requests as quickly as possible. SCHED\_COOP does not need nice priorities, but we assign them to consider the noise caused by other system processes.

\begin{figure}
    \centering
    \includegraphics[width=1\linewidth]{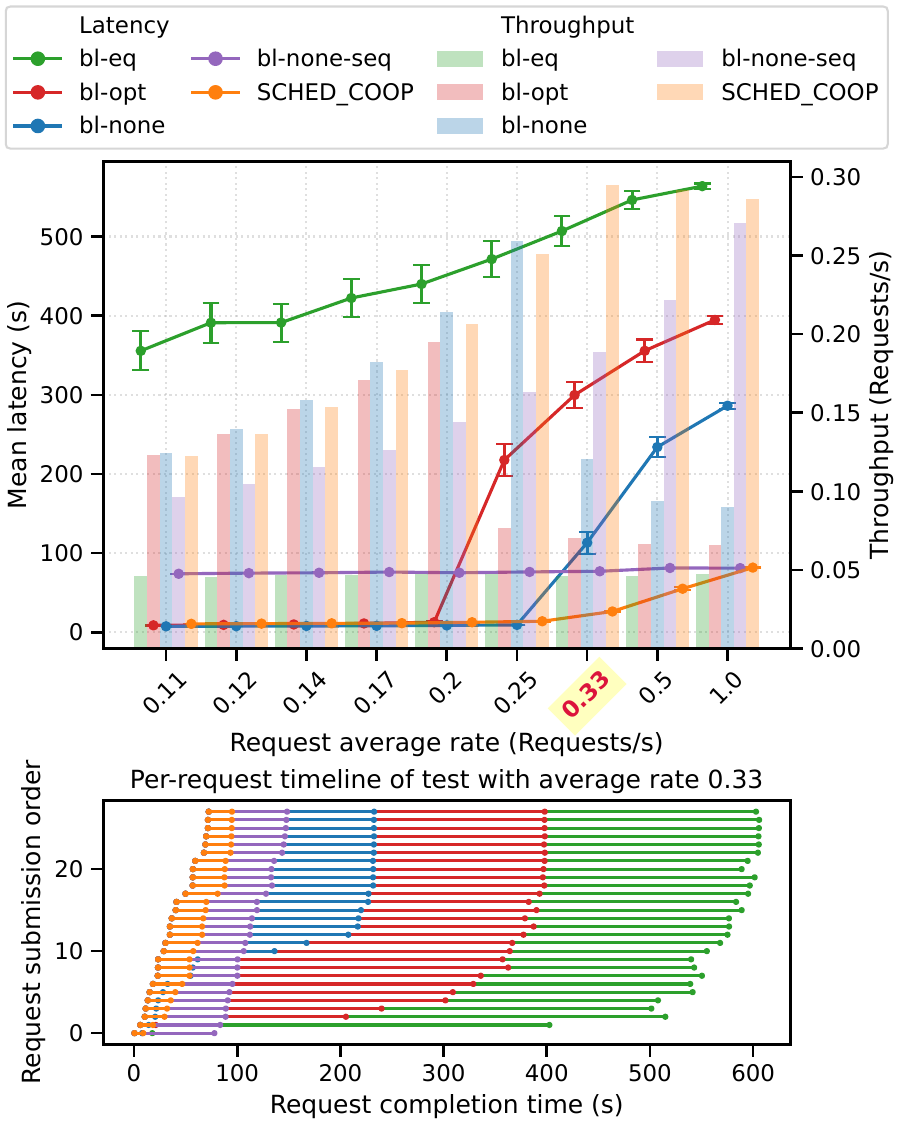}
    \caption{Evaluation of the agentic microservices benchmark. The top figure shows latency and throughput as the frequency of requests increases. The bottom figure shows the detailed per-request start and end time of the test case with rate 0.33. }
    \Description{Multi-panel figure comparing request/response performance under four policies (labeled in the legend as bl-eq, bl-opt, bl-none, and SCHED\_COOP). The top panel (a) show two y axis. The first one is titled "Mean Latency (s)", is a line chart with the x-axis "Request average rate (Requests/s)". Each experiment appears as a separate curve showing how mean latency changes as the offered load increases (x-axis ticks run from low to high request rates). The other y-axis is titled "Throughput", and it is a line chart with the same x-axis (request average rate), plotting one curve per experiment to show how achieved throughput scales with offered load. This panel shows how SCHED\_COOP outperforms the other versions in both throughput and latency, particularly when the request rate is high. The baseline bl-eq is the worst in all cases. bl-opt and bl-none perform well, but for the highest frequency ranges. bl-none obtains a slightly better throughput than SCHED\_COOP for lower frequency ranges. The bottom panel, titled "Request timelines", shows the per-request detail of the case with 0.33 request rate. It shows individual request start-end times in seconds. The figure shows how bl-eq take most of the time to finish, followed by bl-opt. bl-none works much better for the first requests, but the lasts requests consume much more time. Instead, SCHED\_COOP requests finish early even for the lasts requests sent.}
    \label{fig:ms}
\end{figure}

Figure~\ref{fig:ms} (top) shows the throughput and latency obtained for SCHED\_COOP and the three Baselines. As expected, the static partition in \texttt{bl-eq} obtains the worst results for both throughput and latency because of the load imbalance among partitions. \texttt{bl-opt} fixes the imbalance to some extent; although, as the rate increases, the LLaMA server becomes a bottleneck and stalls the overall backend. \texttt{bl-none} gives full control to the Linux scheduler and obtains remarkable results although it eventually collapses because of synchronization noise worsened by oversubscription. \texttt{bl-none-seq} achieves stable latencies across all request rates and high throughput at the highest rates due to the absence of oversubscription noise. However, since servers lack OpenMP parallelism, individual inference times are higher---especially at lower request rates where most cores remain idle. Finally, \texttt{SCHED\_COOP} achieves both low latency and sustained throughput for the whole range of rates. Figure~\ref{fig:ms} (bottom) shows per-request latencies for the run with 0.33 average rate, where \texttt{bl-none} collapses. As ongoing requests accumulate and oversubscription increases, the Linux fair scheduler keeps assigning CPU time evenly to all threads. This causes the requests to progress evenly, including most OpenMP busy-wait barriers and their associated noise, which remain constant until all requests finish nearly simultaneously. Instead, SCHED\_COOP can process requests faster, finish them earlier, and prevent the noise generated by an excessive accumulation of yielding threads, achieving up to 2.4x with respect to \texttt{bl-none}.

As a side note, it would be reasonable to think that non-preemptive schedulers could favor throughput over latency. However, in this scenario, the constant synchronizations of the BLAS libraries act as scheduling points that allow all participating processes to progress despite the lack of involuntary preemptions.

\subsection{LAMMPS and DeePMD-kit  Case Study}
\label{sec:lmp}

This section studies the LAMMPS MD simulator coupled with DeePMD-kit, a deep-learning software that models interatomic potential energies and force fields. Classic quantum mechanical calculations, such as density functional theory (DFT), provide highly accurate results but are computationally prohibitive for large-scale simulations. Specialized AI models trained on data from these first-principles methods can achieve comparable accuracy at a fraction of the computational cost~\cite{mdia,dmd1}. LAMMPS can use MPI to decompose the physical simulation domain into spatial regions, where each MPI rank computes forces and energies for atoms within its assigned region and exchanges boundary data with neighboring ranks. Within each rank, DeePMD-kit runs inferences with PyTorch, which internally leverages an OpenMP-parallelized OpenBLAS for its computational kernels.

The particular atomic spatial distribution and MPI partitioning of a simulation might become imbalanced, as moving particles cross inter-rank boundaries and redistribute each rank workload dynamically at runtime. Hence, it is not trivial to efficiently distribute processes and threads into cores to maximizing resource usage. However, it is a common technique to execute multiple independent simulations or ensembles to explore different paths~\cite{ens1,ens2}, which are usually run one after the other or in different jobs. However, ensembles might add another level of parallelism that we can leverage to improve performance by filling the gaps of one simulation with another. However, at the same time, this further exacerbates the risk of scheduling interference between ensembles, processes, and threads, while under heavy computation phases. 


We analyze multiple execution and oversubscription scenarios of a two LAMMPS ensemble setup with hybrid MPI and OpenMP parallelization. In Exclusive, ensembles run one after the other, use 56 MPI processes with 2 OpenMP threads each, and are pinned to disjoin core sets. The Co-execution scenarios are similar, but the ensembles run in parallel. In Co-location, we avoid oversubscription by reducing the number of MPI processes to 28 (we keep 2 OpenMP threads per rank), and we pin each rank to a disjoint core set. In SCHED\_COOP the ensembles are managed by our scheduler and use 56 MPI and 2 OpenMP each. All configurations but Exclusive feature two variations: In the ``socket'' variant the processes of each ensemble are distributed equally among the two sockets of the machine (half in one socket, and half in the other). Instead, in the ``node'' variant, each ensemble is confined in a different socket. 

We trained a standard DeePMD-kit \emph{se\_e2\_a} model with a descriptor size of [25, 50, 100], a fitting network of [240, 240, 240], and 6\,\text{\AA} cutoff. The simulation creates 20\,K $CH_4$ molecules (5 atoms each, 100\,K atoms in total) created in a box of $292.4\times292.4\times292.4$\,\text{\AA} and distributed in a set of 14 interleaved dense and sparse regions along the x-axis, where dense regions contains 90\% of the atoms, while sparse regions the remaining 10\%. The 3D simulation box is split between ranks along the x-axis, creating an unbalanced environment. We keep a low temperature of 5 degrees Kelvin to keep atom velocities low. We run each simulation for 100 steps. 


\begin{figure*}[htbp]
    \centering
    \begin{subfigure}{\textwidth}
        \centering
        \includegraphics[width=\textwidth]{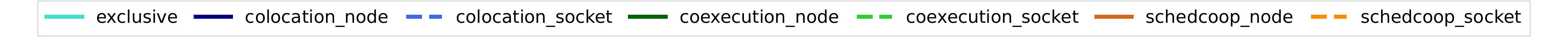}
        \label{fig:lmp_legend}
    \end{subfigure}
    
    \vspace{-0.4cm}  
    \begin{subfigure}{0.29\textwidth}
        \centering
        \raisebox{0.51cm} {\includegraphics[width=1\textwidth]{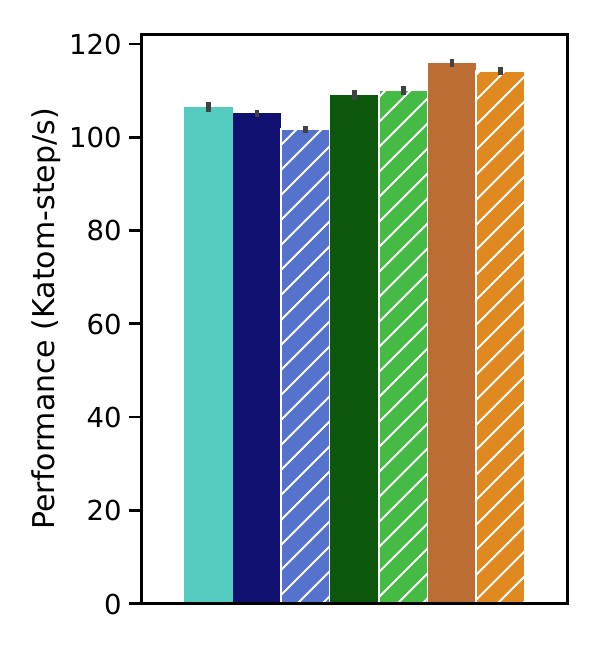}}
        \caption{}
        \label{fig:lmp}
    \end{subfigure}
    \hfill
    \begin{subfigure}{0.7\textwidth}
        \centering
        \includegraphics[width=1\textwidth]{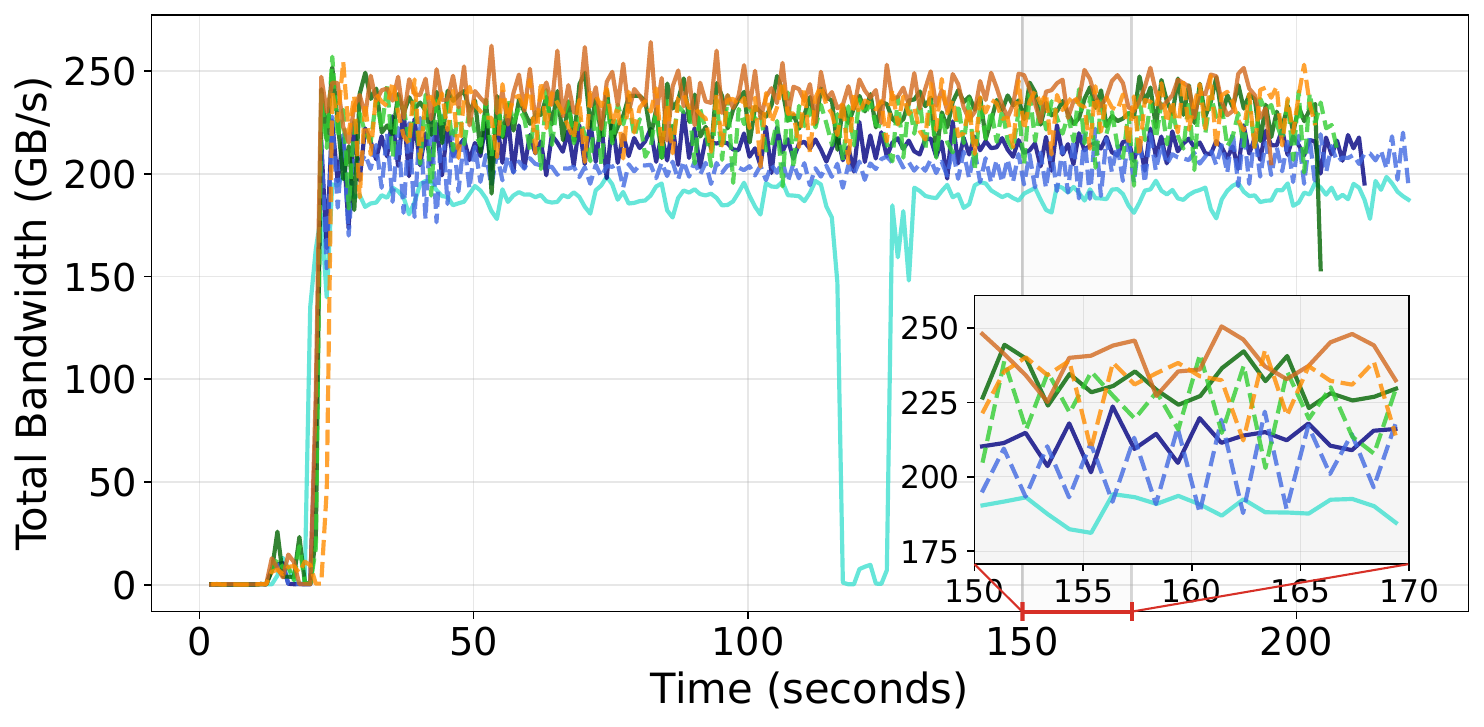}
        \caption{}
        \label{fig:lmp_bw}
    \end{subfigure}
    \caption{LAMMPS and DeePMD-Kit evaluation with two ensembles for several configurations. All ensembles use 56 MPI and 2 OpenMP each, except co-location versions that limit MPI ranks to 28 in order to avoid oversubscription. Figure a) shows performance. Figure b) shows measured total main memory bandwidth for both reads and writes. The inner plot in Figure b) shows a zoomed area of 20s starting at second 150. Average bandwidth in GB/s per scenario is: Exclusive 165.36, colocation\_node 193.09, colocation\_socket 185.80, coexecution\_node 208.17, coexecution\_socket 201.99, schedcoop\_node 214.78, schedcoop\_socket 205.64.}
    \label{fig:combined}
\end{figure*}

Figure~\ref{fig:lmp} shows the evaluation results for all configurations, and Figure~\ref{fig:lmp_bw} compares the total memory bandwidth (reads and writes) observed in the evaluation node during an execution of each configuration. In the Exclusive mode, individual ensembles attain up to 106\,Katom-step/s while in the other configurations, at most 60\,Katoms-setp/s is achieved. That is because there is no interference with other runtimes and each rank can use the full memory bandwidth. However, as shown in Figure~\ref{fig:lmp_bw}, a single ensemble is uncapable of leveraging the whole bandwidth at any point. Moreover, the initialization of each ensemble is sequential (shown as the initial and middle bandwidth valleys), and its cost must be paid per each ensemble; in the other scenarios, the ensemble initialization can be done in parallel, more parallelism is available to exploit the machine resources, and the aggregated Katom-step/s of all ensembles surpass Exclusive.

Still, both co-location scenarios suffer from heavy load imbalance which drives half the caches and cores underused. Besides, the amount of data needed to be kept in the caches with respect to Exclusive doubles because both ensembles are running in parallel. The socket variant gets the worst performance because it also needs to pay the cost of inter-processor communications.

The co-execution scenarios can exploit all cores and caches because 1) each ensemble exposes in total as many threads as cores, and 2) threads can migrate between cores. However, at the same time, the effect of oversubscription and busy wait barriers of both MPICH and OpenBLAS limits their benefits.

Finally, the SCHED\_COOP variants present none of the previous limitations but all their advantages: Running both ensembles in parallel enables more parallel workload to compensate the imbalance, threads can migrate to utilize all resources, and busy waiting is no longer stalling cores. In consequence, SCHED\_COOP attains both the highest Katoms-setp/s and memory bandwidth usage, for up to 4\% speedup compared with the coexecution socket scenario. Still, because USF does not provide support for I/O syscalls, most blocking MPI communications stall cores until they complete. In the future, we will study the integration of these operations to further improve these cases.

\subsection{Takeaways}

From our evaluation, we have learned the following lessons: (1) Applications that combine multiple runtimes are generally complex to tune and debug. Under oversubscription, these applications obtain huge gains from minor code adaptations to avoid busy-wait barriers. (2) Application coexecution is advantageous to maximize resource usage when the involved applications exhibit different degrees of parallelism, but its potential is quickly undermined by the negative effects of preemptive schedulers under oversubscription. (3) SCHED\_COOP can manage both scenarios and, unlike similar approaches, it works transparently, requires no elevated privileges, and is fully compatible with TLS. (4) USF policies are compatible with any application that relies on standard synchronization APIs provided by glibc, but SCHED\_COOP is particularly useful for oversubscribed scenarios with reasonable granularity settings. (5) Cooperative user-space schedulers can harness the potential in oversubscription scenarios of modern applications, previously avoided due to performance degradation. (6) Its main caveat is custom busy-wait barriers. However, we have shown that they are simple to identify and fix.

\section{Related Work}
\label{sec:relwork}

Several works have studied the challenges of coordinating both runtime and kernel schedulers~\cite{omp_imbalance1, omp_imbalance2}. Such challenge grows even larger when multiple different runtimes participate within a single process and oversubscribe the system~\cite{omp_oversub}, such as Python and its high modularity~\cite{python_composition}. These approaches typically cap thread counts, enforce a single runtime, or extend runtimes to coordinate with each other.


It is possible to exploit application imbalance by coexecuting multiple processes on the same node to fill the gaps of one process with another. This has been applied in cloud elasticity~\cite{oversub_cloud}, addressing issues like OS scalability and busy-waiting, though without considering preemptions in critical regions---particularly problematic for parallel runtimes. In HPC, oversubscribing MPI processes can help overlap communication with computation~\cite{oversub_mpi}, but managing multiple runtimes makes this harder. Other work~\cite{oversub_smp} showed that limited oversubscription benefits coarse-grained workloads, yet modern many-core systems are more vulnerable to interrupted synchronizations under system-level preemptive schedulers~\cite{lhp,scalcollaps}. A common alternative is resource partitioning: Arachne~\cite{arachne} and DLB~\cite{dlb} dynamically assign cores to processes at runtime, achieving good results but requiring substantial application or runtime changes to use their APIs.

Recently, there is renewed interest in coexecuting applications as HPC workloads use external Artificial Intelligence (AI) services running as separate processes~\cite{ai_hpc} (AI-out-HPC). However, finding a unified solution to coordinate all involved components and disciplines is challenging. Instead, tackling the problem at the system-level scheduler has the advantage of providing a transparent solution with the potential of greatly simplifying the integration of both worlds.


In that regard, the Linux community has advanced user-space scheduling with the Extensible Scheduler Class (\texttt{sched\_ext})~\cite{sched_ext}, which allows schedulers to be implemented as eBPF~\cite{ebpf} programs. However, decisions are limited to predefined kernel hooks, eBPF code cannot use floating point and requires compile-time loop bounds, and user privileges are needed to load programs. Consequently, Google’s ghOSt~\cite{ghost} and the F4 framework~\cite{f4} propose user-space integrated schedulers without eBPF restrictions, but their reliance on extensive kernel modifications hinders adoption in Linux. Syrup~\cite{syrup} has similar limitations because it relies on ghOSt or eBPF. Purely cooperative user-space schedulers are commonly used to manage User-Level Threads (ULTs)~\cite{argobots,bolt,coro}, but their lack of TLS support restricts their applicability to generic workloads without prior adaptation.


To the best of our knowledge, USF is the first seamless user-space scheduler framework that supports multiple processes and that works within existing Linux environments through glibc extensions. While its SCHED\_COOP policy addresses the fundamental problems of LHP and LWP in oversubscribed environments while keeping TLS support.

\section{Conclusions and Future Work}
\label{sec:conclusions}

In this paper, we present USF, a framework to develop seamless user-space scheduling policies, and SCHED\_COOP, its default cooperative policy. Both are entirely implemented in user-space by extending glibc, require no Linux kernel extensions, and support TLS. Our framework allows the development of ad-hoc scheduling policies and provides the means to define how standard blocking APIs behave. It supports any application and runtime that relies on standard glibc's synchronization primitives, although custom busy wait barriers need to be adapted. We have extensively tested SCHED\_COOP with both multi-process and multi-runtime scenarios and showed that it is effective at managing oversubscription, achieving up to 2.4x.

Future work includes adding support for blocking Linux I/O via \texttt{io\_uring} (as in TASIO~\cite{tasio}); extend the set of supported blocking calls to include \texttt{futex}; expose additional glibc APIs to interface with the USF core; explore an in-kernel implementation of SCHED\_COOP and compare it with the glibcv version; evaluate signal-based preemption in USF; and evaluate new custom scheduling policies targeting different workloads such as mixed criticality systems, or interactive workloads.


\section*{Acknowledgments}
This work was partially supported by the Generalitat de Catalunya (contract 2021-SGR-01007) and the Spanish Government through the Severo Ochoa Program (CEX2021-001148-S/MCIN/AEI/10.13039/501100011033), and it is part of the ST4HPC (PID2023-147979NB-C21) research project, funded by MICIU/AEI/10.13039 /501100011033 and by FEDER, UE. This paper is also co-financed by the Barcelona Zettaescale Laboratory, promoted by the Ministry for Digital Transformation and the Civil Service, within the framework of the Recovery, Transformation and Resilience Plan - Funded by the European Union - NextGenerationEU. We thank the administrators of BSC's Jungle cluster for their technical support.

\balance
\bibliography{bibfile}

\end{document}